\documentclass[preprint2]{aastex}
\slugcomment{Submitted to the Astrophysical Journal on July 20, 2004
\\ Accepted by the Astrophysical Journal on September 1, 2006}

\shorttitle{W49\,B: A Cavity Explosion} \shortauthors{Keohane,
Reach, Rho \& Jarrett}

\begin{document}

\title{A Near-Infrared and X-ray Study of W49\,B:  \\
A Wind Cavity Explosion}

 \author{Jonathan W. Keohane\altaffilmark{1}\altaffilmark{2}, William T.
Reach\altaffilmark{2},  Jeonghee Rho\altaffilmark{2} \& Thomas H.
Jarrett\altaffilmark{2}}
\altaffiltext{1}{The Department of Physics and Astronomy, Hampden-Sydney
College, Hampden Sydney, VA 23943-0716}
\altaffiltext{2}{The Spitzer Science Center,
    The California Institute of Technology,
        MS 220-06, Pasadena, CA  91125-0600}

  \email{jkeohane@hsc.edu, reach@ipac.caltech.edu, rho@ipac.caltech.edu,
jarrett@ipac.caltech.edu}

\email{Submitted to the Astrophysical Journal on July 20, 2004,
Accepted September 1, 2006}

\begin{abstract}
We present near-infrared narrow-band images of the supernova remnant
W49\,B, taken with the WIRC instrument on the Hale 200 inch
telescope on Mt.~Palomar. The 1.64\,$\mu$m [Fe\,II] image reveals a
barrel-shaped structure with coaxial rings, which is suggestive of
bipolar wind structures surrounding massive stars.  The 2.12\,$\mu$m
shocked molecular hydrogen image extends 1.9\,pc outside of the
[Fe\,II] emission to the southeast. We also present archival {\em
Chandra} data, which show an X-ray jet-like structure along the axis
of the [Fe\,II] barrel, and flaring at each end. Fitting single
temperature X-ray emission models reveals:  an enhancement of heavy
elements, with particularly high abundances of hot Fe and Ni, and
relatively metal-rich core and jet regions\@.  We interpret these
findings as evidence that W49\,B originated inside a wind-blown
bubble ($R \sim 5\,{\rm pc}$) interior to a dense molecular cloud.
This suggests that W49\,B's progenitor was a supermassive star, that
could significantly shape its surrounding environment. We also
suggest two interpretations for the jet morphology, abundance
variations and molecular hydrogen emission: (1) the explosion may
have been jet-driven and interacting with the molecular cavity
(i.e.\ a Gamma-ray burst); or (2) the explosion could have been a
traditional supernova, with the jet structure being the result of
interactions between the shock and an enriched interstellar cloud.
\end{abstract}

\keywords{supernova remnants --- supernovae: individual (W49\,B) ---
infrared: ISM --- X-rays: ISM --- gamma rays: bursts ---
circumstellar matter ---  shock waves --- ISM: bubbles}

%%%%%%%%%%  INTRODUCTION  %%%%%%%%%%  INTRODUCTION  %%%%%%%%%%  INTRODUCTION
%%%%%%%%%%

\section{Introduction}  \label{intro.sec}

W49\,B (G43.3-0.2) has the highest radio surface-brightness of all
mixed-morphology  supernova remnants (SNRs) in the Galaxy
\citep{pye84,mof94}\@. SNRs exhibiting centrally-filled X-rays
inside an edge-brightened radio shell are referred to as
mixed-morphology \citep{rho98}\@. A number of models have been
proposed for mixed-morphology supernova remnants
\citep{whi91,cox99,she99,che99}; these models were designed to
explain the larger (i.e. older) SNRs by invoking interactions with a
denser-than-average interstellar medium. The very high radio
brightness and the X-ray properties of W49\,B make it a compelling
object to study in detail, because it may be fundamentally different
from other remnants of its class.

High resolution X-ray spectra of W49\,B \citep{hwa00} revealed
elemental abundances enhanced in heavy elements, suggesting that
W49\,B was the product of a Type Ia explosion.  However, a
subsequent study of \ion{H}{1} absorption \citep{bro01} shows that
W49\,B is at the same distance as the star forming region W49A
\citep[11.4\,kpc,][]{gwi92}\@.   Moreover, long wavelength radio
observations \citep{lac01} suggest that W49\,B is absorbed by H$^+$
gas and is in a high pressure ($\sim$$10^{6} \,{\rm cm^{-3} K}$)
region of the Galaxy, again associating the remnant with the W49\,A
complex.  But at this distance, W49\,B would have to have a massive
progenitor, more akin to a Type II supernova.  Thus we have an
observational inconsistency:  W49\,B has enhanced iron abundances
which are characteristic of a type Ia supernova, yet it is located
in a star forming region and more likely a result of a core-collapse
explosion.

Recently, \citet{mic06} completed a study of W49B with XMM-Newton,
where they concluded that the X-ray emission arises in a
high-metallicity collisionally-ionized plasma, with a temperature
gradient from west to east.

In this {\em Letter} we present near-infrared narrow line imaging
(\S\ref{ir.sec}), which we compare to our spectral analysis of {\em
Chandra} archival data (\S\ref{chandra.sec})\@.  In
\S\ref{discussion.sec} we interpret our results as evidence that
W49\,B was created inside a wind-blown bubble within a molecular
cloud.  We also include two very different interpretations of our
observations: (1) in \S\ref{grb.sec} we interpret our observations
as consistent with a jet-driven explosion (i.e.~Gamma-Ray Burst);
and in \S\ref{sn.sec} we interpret our observations as the result of
a traditional supernova explosion inside a complex cavity.

%%%%%%%%%%  ANALYSIS  %%%%%%%%%%  ANALYSIS  %%%%%%%%%%  ANALYSIS  %%%%%%%%%%

\section{Analysis}  \label{analysis.sec}

 \subsection{Near Infrared Observations}  \label{ir.sec}

\begin{deluxetable}{cccccc}[htb]
%\tabletypesize{\scriptsize}
% \rotate
\tablecaption{WIRC Observing Log for W49\,B \label{wirc.tab}}
%\tablewidth{0pt}
\tablehead{\colhead{Filter}&\colhead{Exposure}&\colhead{Pointings}&\colhead{Seeing}&\colhead{Airmass}
& \colhead{Tot.\ Time} } \startdata
 %    Filter          & NxiT            & Pointings & Seeing      & Airmass  &  total tos & Photometri
 H$_2$           & 3 $\times$ 30\,s & 7        & 0.7\arcsec  & 1.34   &  10.5\,min  \\
 K$_{\rm cont}$  & 3 $\times$ 30\,s & 1        & 0.7\arcsec  & 1.49   &  ~1.5\,min  \\
 ${\rm [Fe\,II]}$ & 2 $\times$ 45\,s & 7        & 0.8\arcsec  & 1.16   &  10.5\,min  \\
 K$_{\rm s}$     & 6 $\times$ 10\,s & 12        & 0.7\arcsec  & 1.38   &  12.0\,min
\enddata
\end{deluxetable}

\begin{figure}[htb]
\plotone{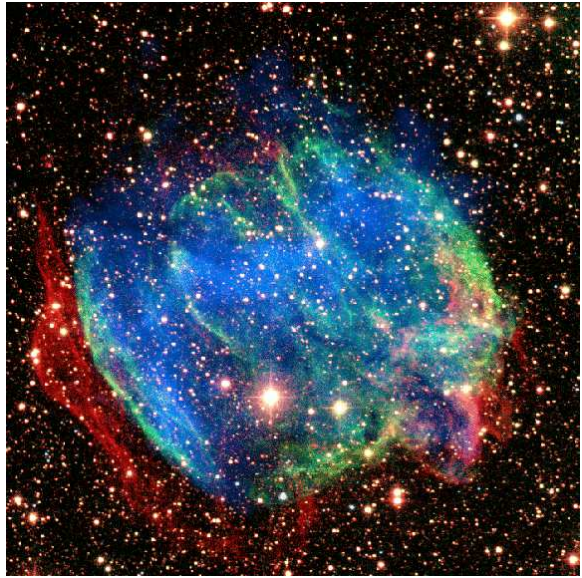}
 \caption{H$_2$ (red), [Fe\,II] (green) and
X-ray (blue) color composite image of W49\,B\@.  The K$_s$ image was
also included as white, in order to produce white foreground stars.}
\label{3color.fig}
\end{figure}

 We observed W49\,B on August 9 and 10, 2003, with the Hale 5\,m telescope
on Mt. Palomar, under clear skies, using the new Wide Field Infrared Camera
\citep[the WIRC, ][]{wil03}\@.  The WIRC is a 2048$\times$2048 Rockwell
Hawaii-II NIR detector mounted at the f/3.3 prime focus, resulting in an
8.7\arcmin\ field of view with 0.25\arcsec\ pixels.

    We observed W49\,B with 4 filters (see Table~\ref{wirc.tab}):  two narrow
line filters and two continuum filters in the 1.6--2.2\,$\mu$m window.  The
purpose of the K$_s$ continuum observation was to search for synchrotron
emission from W49\,B, which was not observed.  With each filter, we alternated
observing on and off our source, with each on-source observation located at a
slightly different position, well sampling our object as well as the nearby
off-source sky.

   Because the WIRC is a new instrument, we developed our own IRAF package to
implement the following procedure independently for each filter: (1)
we subtracted off the median dark image with corresponding frame
times from each image, and corrected for the non-linear response of
the detector; (2) we pixel-by-pixel median averaged our off-source
images, with the highest two values rejected to eliminate stars, to
form a {\em sky image}; (3) we subtracted the {\em sky image} off of
each of our source images; (4) we divided this image by a {\em
standard flat} derived from the linear response of each pixel; (5)
we subtracted a single median off-source background level; (6) we
corrected for cross-field flux bias; (7) we applied a world
coordinate system to our images using the 2MASS point-source
catalog; (8) we flux-calibrated our images using the 2MASS
point-source catalog; (9) we mosaicked our images; (10) we
resubtracted the ambient background level; and (11) we applied a
final calibration using the 2MASS point-source catalog.  The
photometric uncertainty was $\sim$7\% for the spectral line
observations, and $\sim$6\% for K$_s$, and primarily limited by
stellar confusion noise\@.

   The total narrow-band infrared flux densities are 30\,Jy (mag=3.4,
L=1$\times$$10^{37}$\,erg/s) in H$_2$ and 61\,Jy (mag=3.1,
L=3$\times$$10^{37}$\,erg/s) in [Fe\,II]\@.    We used these values
to estimate the mass of Fe$^+$ ions and H$_2$ molecules using
multilevel excitation models and a wide range of
physically-reasonable excitation conditions, including:  all path
lengths shorter than the remnant diameter; all reasonable
temperatures for the observed ionization state; and pressures less
than 1000 times the typical interstellar value \citep[also
see][]{rho01}\@. Thus, the mass of Fe$^+$ must be between 0.2 and 20
$M_\odot$, with the lower and higher masses corresponding to
electron  (temperature, density) of (1000 K, 8000 cm$^{-3}$) and
(700 K, 1600 cm$^{-3}$), respectively.  For H$_2$, the mass is
between 14 and 550 $M_\odot$, with the lower and higher masses
corresponding to H$_2$ (temperature, density) of (2000 K, 2000
cm$^{-3}$) and (1200 K, 3000 cm$^{-3}$), respectively.

  Our calibrated H$_2$ and [Fe\,II] images are shown as red and green
respectively in Fig.~\ref{3color.fig}, while the {\em Chandra} image
is blue.  This figure shows molecular hydrogen emission well-outside
the other emission, especially in the southeast.  The [Fe\,II]
emission appears to be coaxial rings, defining an elongated shell.
The Chandra X-ray emission (see \S\ref{chandra.sec} below) is
interior to the near IR emission, stopping at approximately the same
location as the [Fe\,II]\@. We interpret these data as evidence that
W49\,B was a cavity explosion, as discussed below in
\S\ref{discussion.sec}\@.

\subsection{{\em Chandra} Archival Data} \label{chandra.sec}

\begin{figure}[htb]
\plotone{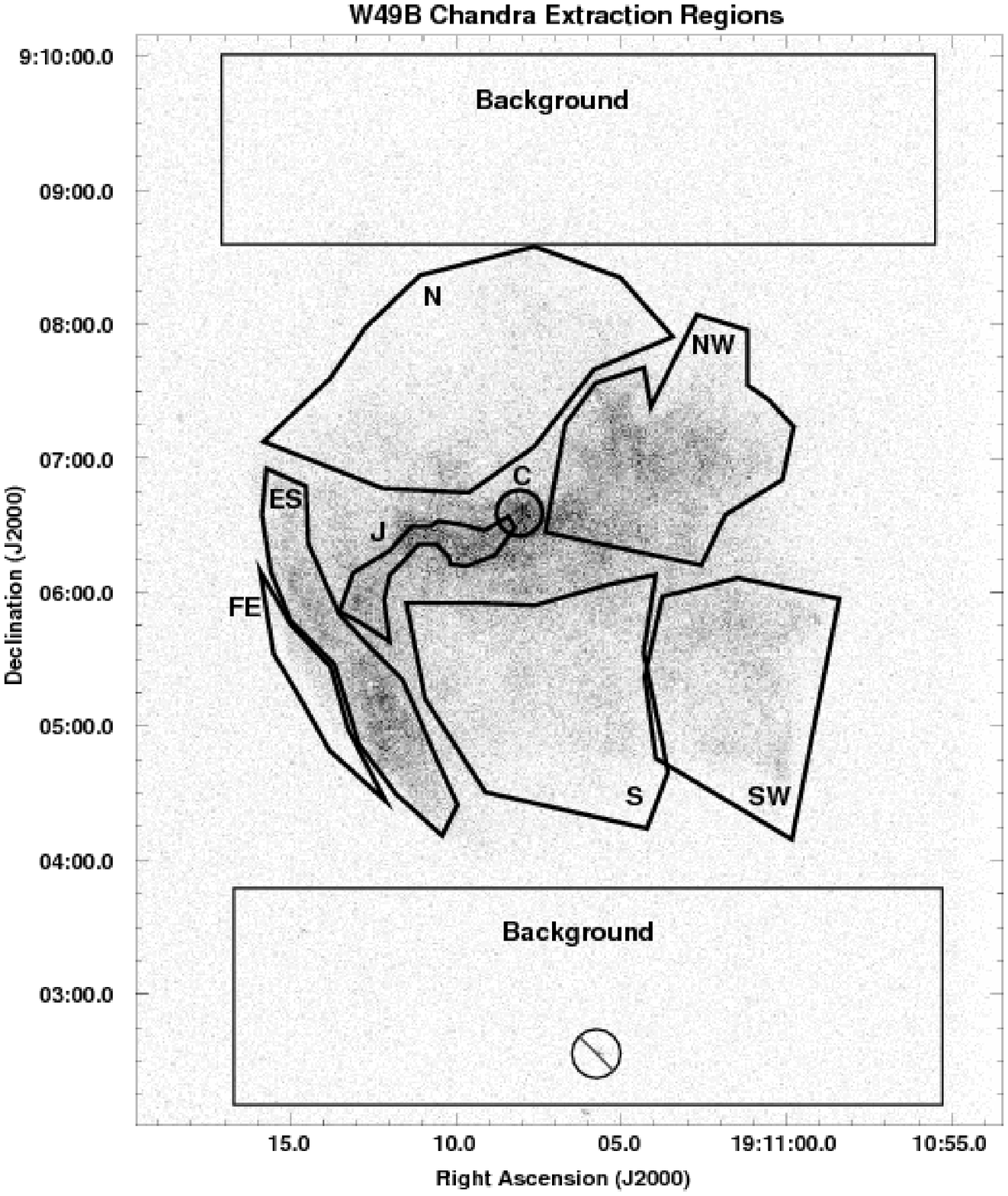} \caption{The Chandra image of W49\,B, with spectral
extraction regions overlaid.  The regions are named from East to
West, : {\em Far East}, {\em East Shell}, {\em Jet}, {\em Center},
{\em North}, {\em South}, {\em Northwest} and {\em Southwest}.  The
approximate central RA and Declination are also shown in
Table~\ref{parameters.tab}} \label{xray_regions.fig}
\end{figure}

\begin{figure}[htb]
\plotone{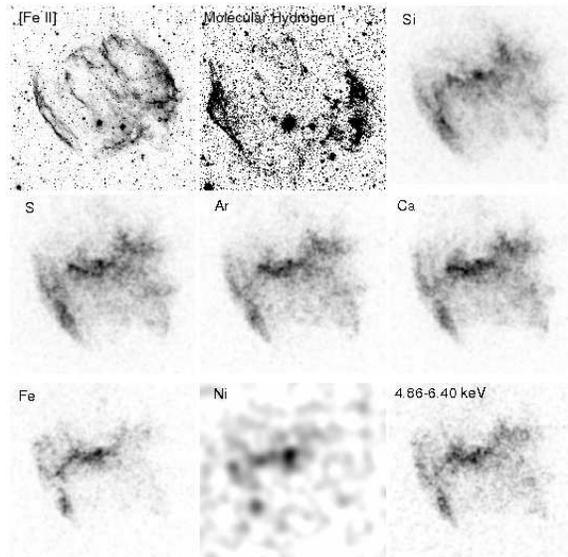} \caption{Chandra spectra emission line images for
Si (1.65-2.1\,keV), S (2.4-2.7\,keV), Ar (3.0-3.35\,keV, Ca
(3.5-4.3\,keV), Fe (6.0-7.2\,keV)and Ni (7.35-8.1\,keV)\@.  The
images were Gaussian-smoothed with $\sigma$=4\,pixels, except for Ni
which was smoothed with $\sigma$=16\,pixels because of its low count
rate. For comparison, we also show the Palomar [Fe II] and H$_2$
images, as well as the {\em Chandra} 4.86-6.40 continuum image.  The
images are scaled linearly to the minimum and maximum surface
brightness.} \label{xray_lines.fig}
\end{figure}

\begin{deluxetable}{cccccccccc}
\tabletypesize{\scriptsize} \rotate \tablecaption{X-ray Spectral Fit
Parameters by Region \label{parameters.tab}} \tablewidth{0pt}
\tablehead{
  &\colhead{Center}&\colhead{Jet}&\colhead{East Shell}&\colhead{Far
East}&\colhead{North}&\colhead{Southwest}&\colhead{South}&\colhead{Northwest}&\colhead{All\tablenotemark{a}}}
\startdata
%                             {Center}            {Jet}&        {East Shell}  {Far East}           {North}&         {Southwest}    {South}  {Northwest}       {All}

\colhead{R.A.} &      19:11:08           & 19:11:11        &
19:11:13  & 19:11:14         & 19:11:09      & 19:11:01    &
19:11:07 &  19:11:03    & --
\\
\colhead{Dec.} &      9:06:40           & 9:60:30          &
9:05:45  & 9:05:30         & 9:07:45       &  9:05:30    &   9:05:15
&   9:07:00    & --
\\
\colhead{N$_{H}$\tablenotemark{b}}&$5.5 \pm0.4 $     &$5.5 \pm 0.2
$&$5.1 \pm 0.2  $&$4.8 \pm 0.5      $&$5.0 \pm 0.2  $&$ 5.8  \pm 0.2
$&$5.3 \pm 0.2 $&$5.2
\pm 0.1 $&$ 5.18 \pm 0.05$\\
\colhead{kT\tablenotemark{c}}&$1.6 \pm0.1 $     &$1.74 \pm 0.06
$&$1.66\pm 0.04 $&$1.4 \pm 0.2      $&$1.57 \pm 0.05$&$ 1.33 \pm
0.05$&$1.32\pm
0.04$&$1.53 \pm 0.04$&$ 1.58 \pm 0.02$\\
\colhead{$\int n_e n_H dV$\tablenotemark{d}}  &$140~^{+60}_{-90}$
&$510 \pm 100 $&$1300 \pm 200   $&$160 \pm 70     $&$1200  \pm
200$&$2800  \pm 300 $&$3500
\pm 300$&$4400 \pm 300 $&$ 16100 \pm 500$\\
\colhead{Si\tablenotemark{e}}&$7~^{+8}_{-3}$    &$4.5~^{+1.4
}_{-0.9}$&$2.8 \pm 0.4  $&$2~^{+2}_{-1}     $&$2.0  \pm 0.4 $&$ 1.1
\pm  0.2$&$1.1 \pm 0.2 $&$1.4
\pm 0.2 $&$ 1.84 \pm 0.08$\\
\colhead{S \tablenotemark{e} }&$4~^{+8}_{-2}$
&$3.6~^{+1.0}_{-0.7} $&$2.8 \pm 0.4  $&$2~^{+2}_{-1}     $&$2.0  \pm
0.3 $&$ 1.2  \pm  0.2$&$1.0 \pm 0.2
$&$1.35 \pm 0.10$&$ 1.83 \pm 0.07$\\
\colhead{Ar\tablenotemark{e}}&$4~^{+8}_{-2}$    &$4 \pm 1
$&$2.5 \pm 0.4  $&$1.1~^{+1}_{-0.7} $&$1.6  \pm 0.4 $&$ 1.1  \pm 0.2
$&$0.8 \pm 0.2 $&$1.3
\pm 0.2 $&$ 1.57 \pm 0.09$\\
\colhead{Ca\tablenotemark{e}}&$5~^{+6}_{-2}$    &$5~^{+2}_{-1}
$&$3.3 \pm 0.6 $&$3 \pm 2        $&$2.7  \pm 0.6 $&$ 2.0  \pm 0.3
$&$1.7 \pm 0.3 $&$2.1  \pm
0.3 $&$ 2.5  \pm 0.2 $\\
\colhead{Fe\tablenotemark{e}}&$12~^{+27}_{-5}$  &$9~^{+3}_{-2}
$&$4.0 \pm 0.6  $&$2~^{+3}_{-1}     $&$3.3  \pm 0.7 $&$ 1.0  \pm
0.2$&$1.3 \pm 0.2 $&$1.5
\pm 0.2 $&$ 2.5  \pm 0.2 $\\
\colhead{Ni\tablenotemark{e}}&$40~^{+110}_{-20}$&$35~^{+14}_{-9}
$&$19  \pm 4   $&$20~^{+27}_{-11}   $&$17    \pm 5  $&$ 9  \pm  3
$&$7   \pm 3   $&$8
\pm 2   $&$ 14  \pm  2  $\\
\colhead{$\chi^2_{\nu}$}&  1.0 &  1.4  &  1.9  &  1.0  & 1.5  & 2.0
&  1.6  &
2.2  & 5.0 \\
\colhead{Mass\tablenotemark{f}}& 0.3 &1.4 &5 &1 &12 &15 &18 &12 & -- \\

\enddata

\tablenotetext{a}{This region encompasses the whole supernova
remnant, and is thus shown for comparison. Note that the single
temperature model is unacceptable for the whole SNR.}
\tablenotetext{b}{The foreground column density is in units of
$10^{22}$cm$^{-2}$} \tablenotetext{c}{The temperature is in units of
keV (i.e. 1.2$\times$10$^7$\,K).} \tablenotetext{d}{$\int n_e n_H
dV$ is in units of pc$^3$ cm$^{-6}$, and assumes a distance of 11.4
kpc.} \tablenotetext{e}{The number per hydrogen relative to solar
values. Abundances of the elements not shown were all set to zero
for this analysis.} \tablenotetext{f}{A very rough estimate of the
mass, in solar masses, assuming a smooth medium. The depth of each
region was assumed to be equal to its width.} \tablecomments{Results
of spectral fitting, using the single temperature equilibrium model
{\sc vmekal}, and the absorption model {\sc wabs}\@.  All errors are
90\% confidence.  The regions used here are shown in
Fig.~\ref{xray_regions.fig}.}
\end{deluxetable}

   The {\em Chandra} X-ray Observatory performed a 55\,ks observation
of W49\,B  in July of 2000 \citep[PI:\ S.\,S.\,Holt;][]{pet00,sta01};
the data became  public a year later.  Morphologically these data show
a double T-shaped structure aligned with the rotation axis of the
progenitor star as discussed above in \S\ref{ir.sec} (see
Fig~\ref{3color.fig})\@.

We obtained screened events files from the {\em Chandra}  Supernova
Remnant Catalog \citep{sew04}\@.   Using standard analysis
techniques\footnote{\texttt{http://cxc.harvard.edu/ciao/}}, we
extracted spectra from the regions shown in
Fig.~\ref{xray_regions.fig}, and made weighted response  functions
using the {\sc acisspec} script.  The data are best-fit using an
absorbed \citep[{\sc wabs}, ][]{mor83} single temperature model
\citep[{\sc  vmekal}, ][]{mew85,mew86,lie95}\@.  The {\sc vmekal}
model does not appear to  be a perfect fit:  it under-predicts the
1.865\,keV line emission from He-like  Si (Si$^{12+}$) and does
include some spectral features between the Ca and Fe  lines
\citep[also seen by ][]{hwa00}\@.  Nevertheless, it is a reasonably
good  fit overall.  We also fit a non-ionization equilibrium model
\citep[i.e.\ {\sc vnei,}][]{ham83}, which fared no better than the
{\sc vmekal} model ($n\,t > 10^{4} \, {\rm cm^{-3} yr}$), implying
that the gas is close to collisional ionization equilibrium\@.

Like \citet{hwa00}, we observe an overall overabundance of heaver
elements.   Moreover, notice from Table~\ref{parameters.tab} the
overall trend toward  higher metallic abundances in the {\em
center}, {\em jet} and {\em eastern  shell}, as opposed to the outer
regions of W49\,B\@.  This same trend can be  seen in images of the
dominant He-like and H-like emission lines (Si, S, Ar, Ca, Fe and
Ni), which are shown in Fig.~\ref{xray_lines.fig}\@.  We also
extracted a 4-6\,keV continuum image, which is morphologically
similar to the  Si, S, Ar and Ca images.

We roughly estimated the total X-ray-emitting mass in each region,
assuming uniform  density and assuming that the depth of each region
was similar to its width.  These mass estimates are also shown in
Table~\ref{parameters.tab}\@. Note that these fits imply a total
X-ray-emitting Fe mass on the rough order of $\slantfrac{1}{10}{\rm
M_{\sun}}$\@.  Also, note that despite the clear  abundance
difference between west and east, the total Fe mass is approximately
symmetrical between the east and the west.

 %%%%%%%%%%  DISCUSSION  %%%%%%%%%%  DISCUSSION  %%%%%%%%%%

\section{Discussion} \label{discussion.sec}

The data presented here imply that W49\,B is the result of an
unusual explosion of a massive or super-massive star.  The
barrel-shaped structure, as seen in the 1.64\,$\mu$m [Fe\,II]
emission, is also seen in the high-pass-filtered radio maps of
\citet{mof94}, suggesting that the location of warm gas is
correlated with high magnetic field.  We interpret these as coaxial
circular rings of enhanced density structures, such as is common in
wind-blown bubbles \citep[e.g. NGC\,6888,][]{par78}\@.  Most
importantly, the [Fe\,II] emission defines the rotation axis of the
progenitor star; which is inclined by 70\arcdeg\ from the line of
sight.  Note that the [Fe\,II] emission arises from much cooler gas
than the X-ray.  We interpret the Fe$^{+}$ gas ([Fe\,II] emission)
to be material from the progenitor's strong winds, while we
interpret the H-like Fe (X-ray Fe lines) to most likely arise from
ejecta (as discussed below in \S\ref{grb.sec})\@.

The H$_2$ emission clearly shows a bow-shock structure in the
southeast, emanating from the point of contact between the X-ray
``jet'' (\S\ref{chandra.sec}) and the shell.  The shock appears to
have traveled a distance of 1.9\,pc inside the molecular gas, the
thickness of the southeastern H$_2$ emission\@. Assuming that the
molecular cloud is uniform density, the shock speed must be less
than 40\,{km/s} in order to not dissociate the H$_2$\@. This
velocity is consistent with an X-ray shock velocity of 1150 km
s$^{-1}$ multiplied by $\sqrt{\slantfrac{\rm n_{X}}{\rm n_{H_2}}}$,
where n$_{\rm X}$ is a density inferred from X-ray gas of 1-3.5
cm$^{-3}$, and $\rm n_{H_2}$ is an H$_2$ gas density of 3000
cm$^{-3}$; note the velocity is inversely proportional to square
root of the density ratio.

The complication with this interpretation is that it would take a
40\,km/s shock 45,000 years to travel the 1.9\,pc that is the
apparent thickness of the H$_2$ shell.  Thus, either: (1) the
remnant actually is 45,000 years old; (2) the thickness of the H$_2$
shell is much less than 1.9\,pc; or (3) the shock is moving much
faster than 40\,km/s.  Each of these possibilities have their own
implications: (1) a 45,000 year old remnant would have required a
larger explosion energy and containment by the cavity to have such a
high current X-ray temperature;  (2) complex projection effects
would need to be invoked to significantly reduce the distance the
shock had to travel in the molecular gas; and (3) a magnetic
precursor could propagate faster without dissociating the molecular
hydrogen.

Elaborating on scenario (3), we can estimate the ionization fraction
in the molecular cloud by assuming that the H$_2$ is excited by a
magnetic precursor propagating at the ion-magnetosonic speed
($v_{\rm ims}$), as in the Cygnus Loop \citep{gra91} and some
Herbig-Haro objects \citep[e.g.\ ][]{HH_object_04}\@.  If we also
assume equipartition between the magnetic and gas pressures, an age
of about 2000 years, and a sound speed of about 3\,km/s
(\S\ref{ir.sec}), the ionization fraction in the molecular cloud
would be $\frac{2 \rho_i}{\rho} \sim 10^{-5}$, because $v_{\rm ims}
\sim \left(\slantfrac{B^2}{4\pi\rho_i}\right)^{\frac{1}{2}}$, where
$\rho_i$ is the ion mass density.

The Chandra (\S\ref{chandra.sec}) and XMM-Newton \citep{mic06} data
show Ni overabundances and concentrated Fe at the {\em center},
which also support an explosion of a massive or super-massive star.
We explore two viable interpretations for this morphology and
chemical structure.  The first interpretation (\S\ref{grb.sec}) is
that W49\,B resulted from a jet-driven explosion producing the {\em
jet}, chemical structure and H$_2$ bow shock structure.  The
alternative interpretation (\S\ref{sn.sec}) is that the explosion
was itself symmetrical, but rather the jet structure is the result
of interactions between the shock and an enriched interstellar
cloud.

\subsection{The Jet-Driven Explosion Interpretation}\label{grb.sec}

A mild consensus has formed regarding the nature of long/soft gamma
ray bursts (GRBs):  they are the result of massive stellar collapse
which produces a highly collimated relativistic blast wave along the
poles of the rotation axis of the core of the progenitor star as it
collapses to form a black hole \citep{woo99, mac01}\@.  This conical
blast wave will continue moving forward until it becomes
semi-relativistic, at which point it will start expanding
perpendicular to the jet \citep{rho97,sar99}\@.  This model has been
applied successfully to the light curves of GRB afterglows,
explaining the sharp steepening of the light curve when the shock
slows down and $\Gamma < \theta_{\rm jet}$, where $\Gamma$ is the
Lorentz factor of the blast wave and $\theta_{\rm jet}$ is the
opening half-angle of the jet, which are believed to be
$\sim$10\arcdeg\ on average \citep{fra01,blo03}\@.  The distance
that the jet travels, the {\em jet-break distance}, depends on the
mass it sweeps-up, but this is typically a few parsecs.  (At a
distance of 11.4\,kpc, the length of the {\em jet} region is about
4\,pc, which would be W49B's observed jet-break distance.)

The frequency of observed GRBs is approximately one per $10^7$ years
per galaxy \citep{sch99}\@.  However, we only observe a fraction of
the explosions $f_b$ due to beaming effects, where $f_b = 1 -
cos(\theta_{\rm jet})$, increasing the rate to about one per 100,000
years per galaxy, assuming the canonical 10\arcdeg\ opening angle.
Thus, depending on the length of time the distinguishing
characteristics remain intact, it is possible that at least one
remnant of a jet-driven explosion could be found in the Milky Way.

These explosions result from the most massive stars, so they should
occur within the molecular clouds which formed them \citep{rei02}\@.
Given that the most massive stars go through phases of high
mass-loss rate (e.g.\ LBV phase) and fast stellar winds (e.g.\
Wolf-Rayet phase), one would also expect their remnants to be
located inside bubbles within molecular clouds
\citep{mir03,che04}\@.  Therefore, the distinguishing
characteristics of a remnant of a jet-driven explosion inside a
massive star should include the following: (1) a double T-shaped
structure, which traces the path of the bipolar jets; (2) a higher
abundance of heavy elements than a typical Type II SNR, because the
jet originates from inside the iron core; (3) a supermassive
progenitor with strong stellar winds; and (4) the nonexistence of a
neutron star. Evidence for these include: (1) the double T-shaped
structure is observed in the X-ray images \citep[See Figures
\ref{3color.fig} and \ref{xray_lines.fig} and the figures
in][]{mic06} and the radio maps \citep{mof94}; (2) {\em Chandra}
(\S\ref{chandra.sec}) and {\em XMM-Newton} \citep{mic06} both
observe an overabundance of Ni and Fe in the {\em center} and {\em
jet} regions; (3) the infrared images show evidence for past stellar
winds interacting with a dense circumstellar medium as would be
expected from a supermassive progenitor; and (4) there is no
evidence for a neutron star in the {\em Chandra} data.

When the star collapses, the resulting twin-jets emerge from the
poles leaving behind material from the stellar core, thus explaining
the enhanced abundances in the {\em center}, {\em jet} and {\em
eastern shell} regions \citep[\S\ref{chandra.sec}, ][]{mic06}\@. The
jet continues until it encounters enough mass to slow to
semi-relativistic speeds (i.e.\ it {\em breaks}), which will happen
at the bubble wall, if not before, because of the large density of
the molecular cloud. This will result in both a transmitted shock
into the dense bubble and a reflected shock back into the cavity.
The morphology of W49\,B (Fig.~\ref{3color.fig}) suggests that the
southeastern jet broke at the bubble wall, while the northwestern
jet broke before hitting the bubble wall.  This interpretation would
explain the clear morphological structure of the southeastern jet,
as well as the more complex northwestern jet. This hypothesis also
explains the relative brightness of the western [Fe\,II] emission
compared to the X-ray, implying cooler gas in the west as observed
here and by \citet{mic06}\@. This is also consistent with the
observed $^{13}$CO map of \citet{sim01}, which shows more molecular
gas to the north and west of W49\,B than the south and east\@.
Moreover, an early western jet-break  would also dilute the ejecta,
explaining the lower abundances in the west compared to the
east---yet there should be about the same Fe mass overall, which is
observed (Tab.\,\ref{parameters.tab})\@.

An issue to consider is the short cooling time of [Fe\,II], which
requires a continuous heating source for the Fe$^+$ gas. Thus,
either the Fe$^+$ shell is currently being shocked, or it is gaining
energy from the adjacent X-ray emitting hot plasma. The thermal
energy currently contained in the X-ray plasma is much greater than
the thermal energy in Fe$^+$ (only about 10$^{45}$ ergs) so this
seems plausible.

\subsection{The Supernova Interpretation}  \label{sn.sec}

An isotropic explosion inside a wind-blown bubble could also produce
these observations, assuming a unique structure of the surrounding
medium. The obvious problem, however, is the clear bipolar structure
in the infrared data images.  A solution may be a strong and ordered
magnetic field, which is consistent with W49\,B's high radio surface
brightness, the radio and [Fe\,II] hoop morphology, and the possible
magnetic precursor in the H$_2$ cloud.

The X-ray jet morphology and abundance variations could possibly be
explained under this model by postulating a metallically enriched
clump of fast moving ejecta which was overtaken by one of the
supernova shocks. The shock would break up the clump, spreading the
material in a line parallel to the shock velocity. Thus the
resulting density enhancement would be elongated radially, and have
an overabundance of heavy elements.

\citet{mic06} compare W49\,B to the SNR G292.08+1.8, which has an
apparently similar X-ray morphology, but has been interpreted to
contain a disk or torus of material viewed edge-on \citep{par04}. We
disfavor this interpretation for W49\,B, because the axis of the
torus would need to be perpendicular to the axis of the bipolar wind
structures presented here, so they could not have been caused by the
same progenitor star.

Another advantage of the supernova model is that W49B's abundances
are more consistent with the 25\,M$_\odot$ models of \citet{mae03}
than their 40\,M$_\odot$ hypernova models \citep{mic06}\@.  However,
given the infancy of detailed GRB models in 2003, and the more
important chemical morphology, we believe this argument to be weak.

This interpretation has the advantage that traditional supernova
explosions, inside complex cloud regions, are common well-known
occurrences.  On the other hand, its primary flaw is that it posits
particular cloud configurations to explain particular morphological
structures.

%%%%%%%%%%  CONCLUSION  %%%%%%%%%%  CONCLUSION  %%%%%%%%%%

\section{Conclusion}\label{conclusion.sec}

We have presented evidence that W49\,B resulted from the explosion
of a supermassive star, inside a wind-blown bubble, which is in turn
interior to a dense molecular cloud.  We have also given two
interpretations for its morphological and chemical structure, each
explanation require a set of extreme initial conditions.  One
interpretation assumed that the axially-symmetric structure is
caused by the explosion mechanism, resulting in the conclusion that
W49\,B was a jet-driven explosion akin to the current model of
long/soft gamma-ray bursts. The other interpretation assumed that
the same observations were rather dominated by complex cloud and
magnetic field structures, and that the explosion could have been a
standard isotropic supernova explosion.

Future work should consist of detailed modeling, with a realistic
preexplosion medium, and with jet-driven and isotropic explosions,
to explain all aspects of the W49\,B observations.

Additional observations would also be useful, especially in order to
understand  complete yields of nucleosynthesis with infrared
spectroscopy by complementing the yields of the X-ray gas, and to
independently map out the molecular cloud structure surrounding
W49\,B.

\acknowledgments

We are grateful to L.\,Rudnick who participated in the observing run
and contributed significant insight. We are grateful to J.\,Hester,
T.\,Pannuti and W.\,Tucker for insightful discussions.  We also
thank R.\,Petre for discussions on W49\,B prior to this work.
Support for this work was provided by NASA through LTSA grant
NRA-01-01-LTSA-013 and Chandra award GO3-4070C awarded to J. Rho\@.

\end{document}